\documentclass[twocolumn,showpacs,preprintnumbers,amsmath,amssymb]{revtex4}
\usepackage{graphicx}
 \newcommand\la{\langle}
 \newcommand\ra{\rangle}
 \newcommand\beq{\begin{equation}}
 
 \newcommand\eeq{\end{equation}}
 \newcommand\beqn{\begin{eqnarray}}
 \newcommand\eeqn{\end{eqnarray}}
 \newcommand\GeV{{\rm GeV}}
 
\def\fm{\,\mbox{fm}}
\def\GeV{\,\mbox{GeV}}
\def\TeV{\,\mbox{TeV}}

\def\Pom{{\bf I\!P}}

\def\lsim{\mathrel{\rlap{\lower4pt\hbox{\hskip1pt$\sim$}}
    \raise1pt\hbox{$<$}}}         
\def\gsim{\mathrel{\rlap{\lower4pt\hbox{\hskip1pt$\sim$}}
    \raise1pt\hbox{$>$}}}         
\def\BA{\begin{eqnarray}}
\def\BE{\begin{equation}}
\def\BF{\begin{figure}[htb]}
\def\BT{\begin{table}[htb]}
\def\EA{\end{eqnarray}}
\def\EE{\end{equation}}
\def\EF{\end{figure}}
\def\ET{\end{table}}

\def\la{\langle}
\def\ra{\rangle}

\begin{document}

\title{{\boldmath$J/\Psi$} in high-multiplicity pp collisions: lessons from pA}

\author{B. Z. Kopeliovich$^1$}
\author{H. J. Pirner$^{2}$}
\author{ I. K. Potashnikova$^1$}
\author{K. Reygers$^{3}$}
\author{Iv\'an Schmidt$^{1}$}

\affiliation{\centerline{$^1$Departamento de F\'{\i}sica,
Universidad T\'ecnica Federico Santa Mar\'{\i}a; and}
Centro Cient\'ifico-Tecnol\'ogico de Valpara\'iso;
Casilla 110-V, Valpara\'iso, Chile
\\
{$^{2}$Institute for Theoretical Physics, University of Heidelberg, Germany
}\\
{$^3$Physikalisches Institut, University of Heidelberg, Germany}}

\begin{abstract}
Gluons at small $x$ in high-energy nuclei overlap in the longitudinal direction,
so the nucleus acts as a single source of gluons, like higher Fock components in a single nucleon,
which contribute to inelastic collisions with a high multiplicity of produced hadrons. This similarity helps to make a link between nuclear effects in $pA$ and high-multiplicity $pp$ collisions.
Such a relation is well confirmed by data for the $J/\Psi$ production rate in high-multiplicity $pp$ events
measured recently in the ALICE experiment.
Broadening of  $J/\Psi$ transverse momentum is predicted for high-multiplicity $pp$ collisions.
\end{abstract}



\maketitle

\section{Introduction}

Hadron multiplicities  larger than the mean value in $pp$ collisions
can be reached due to the contribution of higher Fock states in the proton, containing an increased number of gluons. 
Correspondingly, the relative rate of $J/\Psi$ production will be also enhanced, because heavy flavors are produced more abundantly in such gluon rich collisions.

We define the ratios of measured multiplicity to average multiplicity in $pp$ collisions per unit of rapidity as $R$, and differentiate between the general hadron multiplicity ratio $R_h$ and the $J/\Psi$ ratio $R_{J/\Psi}$: 
\beq
R_{h}^{pp}\equiv \frac{dN_{h}^{pp}/dy}{\left\la dN_{h}^{pp}/dy\right\ra}
\label{100}
\eeq
\beq
R_{J/\Psi}^{pp}\equiv \frac{dN_{J/\Psi}^{pp}/dy}{\left\la dN_{J/\Psi}^{pp}/dy\right\ra}.
\label{200}
\eeq
The denominators in (\ref{100}) and (\ref{200}) are the mean charge hadron multiplicity and the mean $J/\Psi$ multiplicity per event averaged over events with different hadron multiplicities.

More gluons participating in collisions with $R_h^{pp}>1$ explain why $R_{J/\Psi}^{pp}$ rises with increasing $R_h$. Of course such fluctuations are rare. The absolute value of the $J/\Psi$ production rate in such rare events might be very low.  
Although qualitatively such a correlation is rather obvious, its quantitative description is far from being trivial.
In this paper we outline and employ a close relation between the $R_{J/\Psi}^{pp}$ -$R_h^{pp}$ correlation in $pp$ and $pA$ collisions.

In a boosted high-energy nucleus the longitudinal distances between the bound nucleons are contracting
with the Lorentz factor $m/E$, while the small-$x$ glue in each nucleon contracts much less, as $m/xE$
\cite{kancheli}.
For instance, in a collision at the c.m. energy $\sqrt{s}$ of LHC at the mid rapidity, 
$x=2\sqrt{\la k_T^2\ra}/\sqrt{s}\sim 10^{-4}$, where $\la k_T^2\ra\approx 0.5\GeV^2$ is the mean transverse momentum squared of gluons. So gluons stick out far from the Lorentz contracted nuclear disc. In the nuclear rest frame the same effect is interpreted as a long lifetime of 
the gluon cloud, which propagates in the longitudinal direction over the distance 
$\sqrt{s}/\la k_T^2\ra$, four orders of magnitude longer than the mean inter-nucleon spacing in nuclei at LHC energies. Therefore all gluons, which overlap in the transverse plane, also overlap longitudinally.
Thus, they can be treated as a single gluon cloud originating from one source with increased  density,
equivalent to higher Fock states in a single nucleon.

We claim that one can emulate the dependence of $R_{J/\Psi}^{pp}$ on $R_h^{pp}$  in high-multiplicity pp collisions by  the analogous correlation in $pA$ collisions. In the latter case, one can use as the numerators in the ratios (\ref{100}) and (\ref{200}) the mean multiplicities of light hadrons and $J/\Psi$  measured in $pA$ collisions, while the denominators remain the same as in (\ref{100})-(\ref{200}), so they are also here
the mean hadron and the $J/\Psi$  multiplicities in $pp$ collisions.

\section{\boldmath High-multiplicity events in $pp$ and $pA$ collisions}

\begin{figure}[htb]
\begin{center}
 \includegraphics[width=6cm]{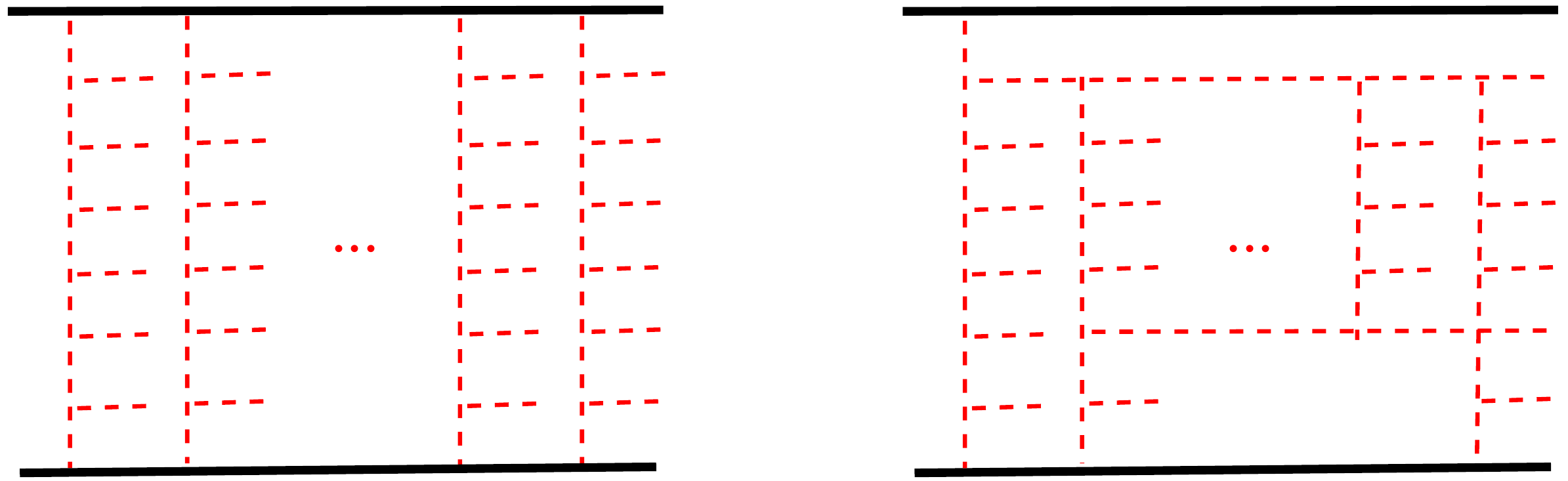} 
 \caption{\label{fig:eikonal} (Color online) Amplitude of multiple production corresponding to $n$ cut pomerons in the eikonal approximation ({\sl left}), and including the Landau-Pomeranchuk coherence effects ({\sl right}).}
  \end{center}
 \end{figure}

Multiplicity distributions in $pp$ and $pA$ collisions at high energies have been studied \cite{ak,karen1,karen2} within Regge phenomenology, based on the Abramovsky-Gribov-Kancheli (AGK) cutting rules \cite{agk}. The simultaneous unitarity cut of elastic  $pp$ amplitude through $n$ pomerons corresponds to the production of $n$ showers of particles, i.e. to a multiplicity $n$ times higher than in one cut pomeron. 
Weight factors for these graphs are usually estimated either within the eikonal model (see Fig.~\ref{fig:eikonal}, left), or in the quasi-eikonal approximation \cite{karen3}, taking into account intermediate diffractive excitations.

Particle production in $pA$ collisions has many similarities to high-multiplicity $pp$ events. The Glauber model, equipped with the AGK rules, corresponds also to the eikonal graphs in Fig.~\ref{fig:eikonal} (left).
However, in pA-collisions the Pomerons are attached to different nucleons in the nuclear target.
For this reason high-multiplicity events in $pA$-collisions are enhanced, because
the weight factors are different from $pp$, and the graph with $n$ cut Pomerons (production of $n$ showers) contains the factor $A^{n/3}/n!$. As a result, the inclusive cross section of particle production acquire no nuclear shadowing \cite{kancheli70,mueller}.
The mean number of produced showers, so called number of collisions, 
\beq
N_\mathrm{coll}\equiv
A\,\frac{\sigma_\mathrm{in}^{pN}}{\sigma_\mathrm{in}^{pA}}.
\label{270}
\eeq
and the mean hadron multiplicity increase as $A^{1/3}$. 
The nuclear ratio, $R_h^{pA} = \la dN_h^{pA}/dy\ra/\la dN_h^{pp}/dy\ra$
is defined as a ratio of the mean hadron multiplicities in $pA$ to $pp$ collisions.
In the Glauber eikonal model it is given by the number of collisions,
\beq
\left(R_h^{pA}\right)_\mathrm{Gl}=N_\mathrm{coll}.
\label{280}
\eeq
However, comparison with data shows that this relation significantly overestimates the hadron multiplicity. The popular parametrization for the nuclear effects,
\beq
R^A_h=1+\beta(N_\mathrm{coll}-1),
\label{300}
\eeq
shows, when fitted to data, that $\beta<1$  \cite{huefner}.

The eikonal Glauber approximation Eq.~(\ref{280}) ignores several corrections. Energy conservation shrinks the allowed energy interval with rising number of cut pomerons \cite{froissaron}. The 
source partons, i.e. the valence and sea quarks, which participate in the multi-pomeron exchange, are distributed in rapidity, and this affects the energy and multiplicity in each cut pomeron. This is taken into account in the quark-gluon string \cite{qgsm} or dual parton \cite{capella} models, which describe quite well multiple hadron production \cite{kaidalov,kaidalov-karen}.

Another source of reduction of multiplicity are coherence effects. The eikonal description relies on the Bethe-Heitler regime of radiation illustrated in Fig.~\ref{fig:eikonal} (left).  However, amplitudes of gluon radiation from inelastic interactions on different nucleons interfere, leading to a damping of the radiation spectrum, known as Landau-Pomeranchuk suppression, or gluon shadowing. 
The related radiation pattern is illustrated in Fig.~\ref{fig:eikonal} (right).
Maximal suppression occurs when the gluon density \cite{glr,al} saturates,
which leads to a modification of the transverse momentum distribution of gluons called color glass condensate \cite{mv}.

A comprehensive analysis of data  \cite{busza} from fixed-target experiments led to $\beta=0.59\pm0.01$. The recent analysis of data in \cite{phobos}, at $\sqrt{s}<200\GeV$, found good agreement 
with the simple behavior $R^A_h={1\over2}\,N_\mathrm{part}$, where the number of participants for $pA$ collisions is $N_\mathrm{part}^{pA}=N_\mathrm{coll}^{pA}+1$. This relation is equivalent to Eq.~(\ref{300}) with $\beta=0.5$. 
While this value of $\beta$ in Eq.~(\ref{300}) underestimates data for $R_h^{pA}$ by about one standard deviation, a larger value $\beta=0.65$ leads to an overestimation by a similar magnitude. For further calculations we treat the interval $0.5<\beta<0.65$ as a measure of experimental uncertainty.
This interval agrees with the multiplicity measured recently at $\sqrt{s}=5\TeV$ in \cite{alice-mult}, $dN_h^{pA}/dy=17.24\pm 0.66$, which leads to $\beta\approx0.55$.

Finally we also want to relate $R^A_h$ directly to the nuclear mass number $A$. We evaluate the $A$ dependence of $N_\mathrm{coll}$ in Eq.~(\ref{270}),
using as a nuclear radius $R=r_0 A^{1/3}$ with $r_0\approx 1.12\fm$ and equating the nuclear inelastic cross section with the geometrical one. Then we obtain for heavy nuclei     
\beq
N_\mathrm{coll}\approx \frac{\sigma_\mathrm{in}^{pp}}{\pi r_0^2}\,A^{1/3}.
\label{360}
\eeq
Although $N_\mathrm{coll}$ can be calculated much more precisely, even including Gribov corrections \cite{mine,kps-heraB} and $NN$ correlations in nuclei \cite{ciofi}, the accuracy of calculation in (\ref{360}) hardly affects the final result (see below).

\section{\boldmath Correlation between the multiplicity and $J/\Psi$ rate  on nuclei}

Both the $J/\Psi$ production rate and the mean multiplicity of light hadrons in $pA$ collisions rise with $A$.
Here we attempt to  relate them directly. 

The mechanisms of the nuclear effects in $J/\Psi$ production has been debated since the first accurate measurements at SPS in the NA3 experiment \cite{na3}. Besides the usual nuclear enhancement of hard processes by the factor $N_\mathrm{coll}$, the production rate of $J/\Psi$ exposes a significant suppression, especially at forward rapidities. Depending on the collision energy, the mechanisms of suppression can include energy loss and break up of the $\bar cc$ dipoles, higher twist shadowing of charm quarks and leading twist gluon shadowing (see review \cite{puzzles}).
No consensus has been reached so far with respect to the mechanisms of $J/\Psi$ production, either in $pA$ or even in $pp$ collisions. Therefore we prefer to rely on data here.

The results of the fixed-target experiments NA3 \cite{na3} and E866 \cite{e866} on different nuclei show that the cross section of $J/\Psi$ production on nuclei  
can be parametrized  as $A^\alpha$, so according to Eq.~(\ref{200})
\beq
R_{J/\Psi}^{pA} \equiv \frac{\left\la dN_{J/\Psi}^{pA}/dy\right\ra}
{\left\la dN_{J/\Psi}^{pp}/dy\right\ra}=
\frac{N_\mathrm{coll}}{A}\,
\frac{d\sigma_{J/\Psi}^{pA}/dy}{d\sigma_{J/\Psi}^{pp}/dy}=
N_\mathrm{coll}\,A^{\alpha-1},
\label{340}
\eeq
where $\alpha$ depends on $s$ and $y$. The results of the E866 experiment, which have the best accuracy, give  $\alpha=0.95$ at $y=0$. NA3 data \cite{na3} at lower energy and PHENIX data \cite{phenix-psi} at $\sqrt{s}=200\GeV$  agree with this value. So far the ALICE experiment  has measured $J/\Psi$ 
only at forward  rapidities, with a similar nuclear suppression \cite{alice-psi}.

While the first factor, $N_\mathrm{coll}$ in (\ref{340}), is directly related to $R_h^{pA}$ by Eq.~(\ref{300}),
the second factor, $A^{\alpha-1}$, can be evaluated with Eq.~(\ref{360}).
Then we arrive at,
\beq
R_{J/\Psi}^{pA}= \left(1+\frac{R_h^{pA}-1}{\beta}\right)^{3\alpha-2}
\left(\frac{\sigma_\mathrm{in}^{pp}}{\pi r_0^2}\right)^{3(1-\alpha)}.
\label{380}
\eeq
Since the exponent in the last factor is very small, the accuracy of calculations in Eq.~(\ref{360})
does not affect much the result (\ref{380}).

Notice that as long as $\alpha$ does not vary in a wide energy range, as was discussed above, the relation (\ref{380}) is nearly energy independent. Indeed, the first factor apparently has no energy dependent ingredients, only the second factor contains the slowly rising $\sigma_\mathrm{in}^{pp}\propto s^{0.1}$, which results in an extremely weak overall energy dependence $\propto s^{0.015}$.

The exponent $\alpha$ is known to vary with rapidity. It drops significantly at large Feynman $x_F$
in the fixed-target experiments \cite{na3,e866}. Moreover, data agree that $\alpha$ scales with $x_F$.
However, data taken so far at RHIC and the LHC correspond to very small $x_F$ and do not show any clear
dependence of $\alpha$ on $y$. Although theoretical models predict a falling behavior of $\alpha$ with $y$,
we prefer here to rely on data and provide numerical predictions only at mid-rapidity. 

Eq.~(\ref{380}) is the final relation between the multiplicities of $J/\Psi$ and light hadrons in $pA$ collisions, which we are going to apply to high-multiplicity $pp$ collisions.

\section{\boldmath Bridging $pA$ and high-multiplicity $pp$ collisions}

Before one links  multi particle processes in $pp$ and $pA$ collisions
one may argue that there exists an  essential difference.  Whereas high-multiplicity $pp$ collisions necessitate symmetric higher Fock components in both protons, $pA$ collisions look asymmetric.
Although the small-$x$ gluon cloud from several longitudinally overlapping nucleons in a high-energy nucleus acts like a higher Fock state in a single nucleon, the proton on the other side may still be in an averaged Fock state.  This argument, however, is not correct, because the weight factors for different Fock states in the proton depend on the way they are probed. The scale evolution of the parton distribution function represents a well known example: the gluon density in the proton at small $x$ steeply rises with the $Q^2$ of the photon probing it.
Therefore the mean parton configuration in the proton also drifts to higher Fock components when probed by a collision with a nucleus (see Ref. \cite{boosting}).

\begin{figure}[htb]
\begin{center}
 \includegraphics[width=4cm]{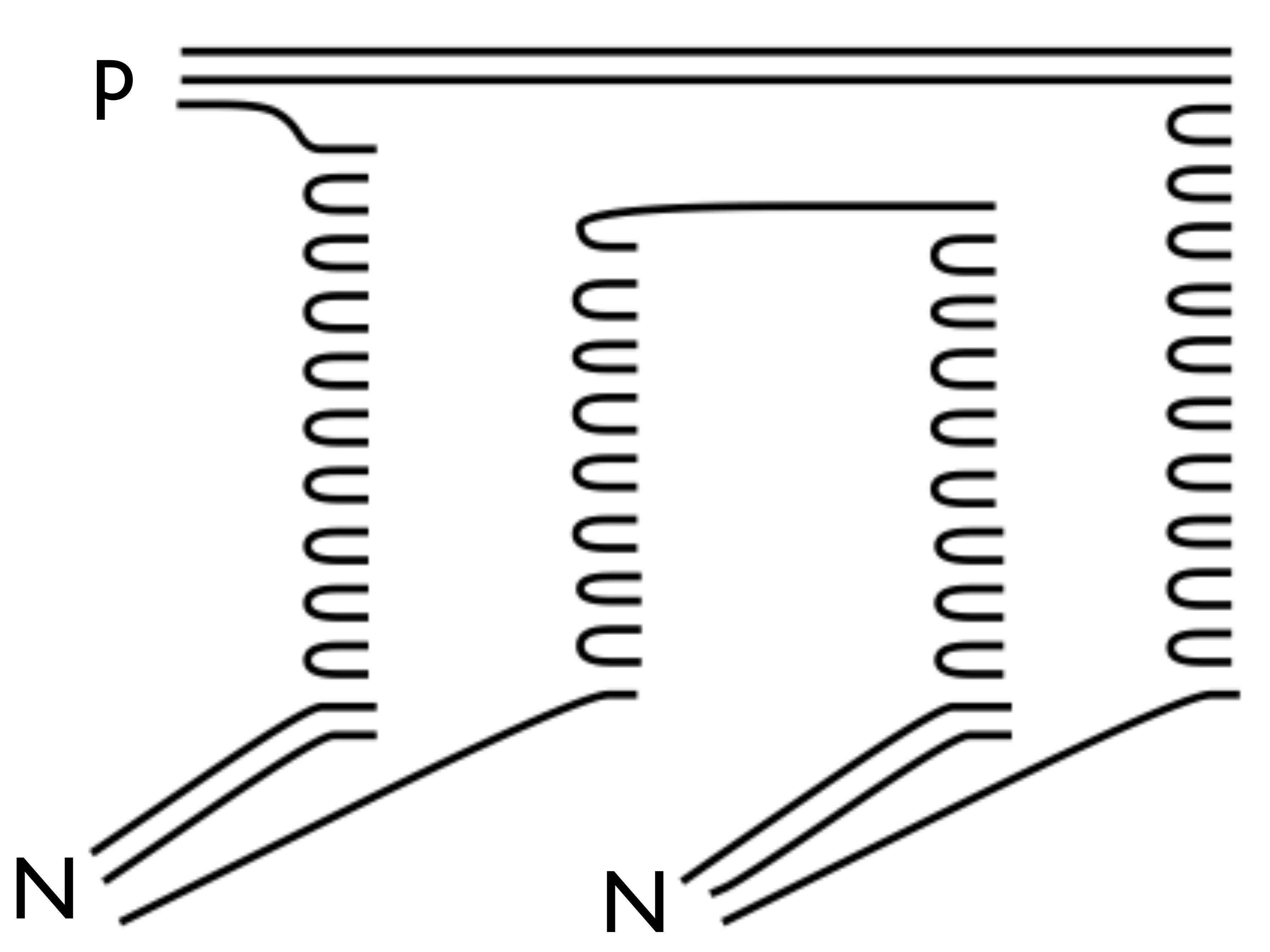} 
 \caption{\label{fig:capella} Double scattering of a proton in a nucleus in the dual topological description.}
  \end{center}
 \end{figure}

This idea is explicitly realized in the quark-gluon string \cite{qgsm} or dual \cite{capella} models.
Multiple inelastic interactions in a $pA$ collision are not sequential, but occur "in parallel", i.e.
form a multi-sheet topology. Otherwise, several pomerons could not undergo a simultaneous unitarity cut
as is requested by the AGK cutting rules. 
An example of a double interaction of the proton with two bound nucleons \cite{cap-kop} is depicted in Fig.~\ref{fig:capella}.
One can see in Fig.~\ref{fig:capella} that the proton undergoing multiple interactions has the same number of endpoints of strings as both bound nucleons together, namely two color triplets and two anti-triplets. Therefore the color content inside the proton and the nucleus for a $p$-$2N$ collision looks symmetric. 

Notice that semi-enhanced fan-type reggeon graphs make the rapidity dependence of the multiplicity distribution asymmetric \cite{koplik}. However, these graphs are large and the asymmetry is significant only in the nuclear fragmentation region. In the central rapidity region at high energies, the fan diagrams are suppressed by the smallness of the triple-pomeron coupling, which originates from the smallness of gluonic dipoles within the QCD description \cite{spots}.

Another source of a possible distinction between high-multiplicity $pp$ and $pA$ collisions is the difference in the impact parameter pattern of multiple interactions. 
It has been known since the early era of Regge theory  that multi-shower particle production is characterized by  smaller impact parameters of collision than in the production of a single shower, and this is a direct consequence of the AGK cutting rules.
Indeed, the mean impact parameter squared for a single pomeron exchange is $\la b^2\ra_{\Pom}=2B_{\Pom}(s)$,
where $B_{\Pom}(s)=B_0+2\alpha^\prime_{\Pom}\ln(s/s_0)$ is the standard Regge parametrization for the energy dependent elastic slope \cite{bkk}, $d\sigma/dt\propto \exp\left[B_{\Pom}(s) t\right]$.

The slope for the elastic amplitude with $n$ Pomeron exchange calculated in the eikonal model is $n$ times smaller $B_{n\Pom}(s)=B_{\Pom}(s)/n$. Thus, the events with multiplicity $n$ times higher than the mean value are produced in collisions with impact parameters as small as,
\beq
\la b^2\ra_{n\Pom}={1\over n}\la b^2\ra_{\Pom}.
\label{250}
\eeq
Smallness of the mean impact parameter of a collision means  larger transverse momentum of the parton, 
\beq
\la k_T^2\ra_n= n\,k_0^2,
\label{255}
\eeq
 where $k_0$ is the transverse momentum gained in a single pomeron interaction. 

On the other hand, a parton propagating through a nucleus undergoes multiple interactions with different nucleons,
with impact parameters much larger than in (\ref{250}). Nevertheless, the total transverse momentum gained by the parton is the same as in (\ref{255}), since the single-Pomeron interaction with every nucleon remains the same, $k_0$. Moreover, in high-multiplicity events in $pA$ collisions, where one should convolute multiple-pomeron interactions with separate nucleons with increasing number of collisions,
the result (\ref{255}) remains valid. Indeed, comparison of $\la k_T\ra$ measured in $pp$ and $pA$ collisions at equal hadron multiplicities demonstrate the equality of the mean transverse momenta at not too large multiplicities $R_h^{pA} \lesssim 5$, which is the range of our further calculations.

However, at higher multiplicities $\la k_T\ra$ in $pp$ collisions was found to be considerably higher than in $pA$
\cite{morsch}. 
This remarkable observation clearly shows an onset of a new dynamics at very high $R_h^{pA}$,
related to the existence of two scales in the proton. The semihard scale corresponds to the short-range glue-glue correlation radius \cite{pisa}, or the small size of instantons \cite{shuryak}. Small gluonic spots in the proton \cite{kst2,spots}, which is a minor effect in the elastic $pp$ scattering \cite{k3p,totem}, give a significant and rising contribution at high multiplicities. They are characterized by a much higher transverse momenta of gluons and lead to a steep growth of $\la k_T^2\ra$ at high multiplicities. This is a much smaller effect in $pA$ collisions, which gain high multiplicities mainly by increasing $N_\mathrm{coll}$.

While these two scales affect light hadron production, they have practically no influence on the production of $J/\Psi$, which is characterized by an order of magnitude higher scale. For this reason, we can safely apply the results of $J/\Psi$ production in $pA$ collisions to high-multiplicity events in $pp$ collisions.

\section{\boldmath $J/\Psi$ production in high-multiplicity $pp$ collisions}

Now we are in a position to rely on nuclear effects observed (or calculated) in $J/\Psi$ production in $pA$ collisions, attempting to predict analogous effects in high-multiplicity $pp$ events. 

\subsection{Production rate}

\begin{figure}[htb]
\begin{center}
 \includegraphics[width=8cm]{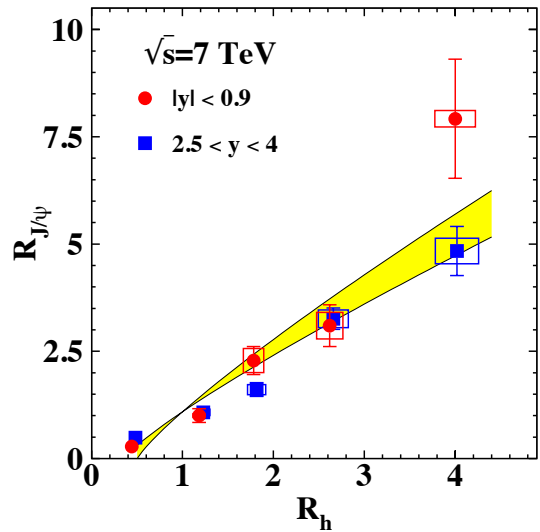} \\[-20mm]
 \caption{\label{fig:data} (Color online) Normalized multiplicity of $J/\Psi$, $R_{J/\Psi}$, vs normalized multiplicity of charged hadrons, $R_h$. Data from ALICE \cite{alice-psi-mult} for $pp$ collisions at $\sqrt{s}=7\TeV$ are plotted as round points (red) for $y<0.9$, and as squares (blue) for $2.5<y<4$.
 The upper and bottom curves show the relation (\ref{380}), predicted based on data for $pA$ collisions at $y=0$, with $\beta=0.5$ and $0.65$ (see Eq.~(\ref{300})), respectively. }
  \end{center}
 \end{figure}

We assume that the relation (\ref{380}), derived for nuclear targets,
can be applied to $J/\Psi$ production in $pp$ collisions.
Within the above mentioned uncertainty in the value of $\beta$, the relation Eq.~(\ref{380})  is plotted in Fig.~\ref{fig:data} as the yellow strip. We can now test our hypothesis that the dependence of $R_{J/\Psi}^{pp}$ on  $R_{h }^{pp}$
in $pp$ collisions is the same as the dependence of  $R_{J/\Psi}^{pA}$ on $R_{h}^{pA}$ in $pA$ collisions, by comparing  with data for high-multiplicity $pp$ collisions \cite{alice-psi-mult}. We see that the dependence predicted from data on $pA$ collisions agrees well with $pp$ data at mid rapidity.
Notice that although the second factor in (\ref{380}) was calculated approximately in the black disc limit, the result is rather accurate due to smallness of $1-\alpha$.

\subsection{\boldmath $p_T$ broadening of $J/\Psi$}

The analogy between high-multiplicity $pp$ events and $pA$ collisions can be extended further.
It was observed experimentally \cite{e866}, and well understood theoretically \cite{jkt,broad},
that the mean transverse momentum squared of the $J/\Psi$ increases in $pA$ compared to $pp$ collisions. 

As far as the gluon radiation time exceeds the nuclear size, the radiation process does not resolve between a single and multiple interactions, but is sensitive only to the total kick to the scattering color charge. For instance, if a parton gets the same momentum transfer interacting with a proton or with a nuclear target, the radiation of gluons with large $l_c\gg R_A$ should be the same. This means that multiple interactions, either in high-multiplicity $pp$ interactions or in $pA$ collisions, affect the $p_T$ distribution of produced $J/\Psi$ 
similarly, leading in both case to broadening defined as,
\beq
\Delta p_T^2\equiv\la p_T^2\ra_{R_h^{pp}>1}-\la p_T^2\ra_{R_h^{pp}=1}.
\label{390}
\eeq

The rise of the mean transverse momentum squared, for a gluon propagating through a nucleus at impact parameter $b,$ was calculated in \cite{broad,jkt,dhk} as,
\beq
\Delta p_T^2(b) = {9\over2}\,C(E)\,T_A(b),
\label{700}
\eeq
where $E$ is the gluon energy in the nuclear rest frame; the nuclear thickness function, $T_A(b)=\int_{-\infty}^\infty dz\rho_A(b,z)$ is given by the integral of the nuclear density $\rho_A$ along the parton trajectory.

The coefficient $C(s)$ controls the behavior of the universal dipole cross section at  small dipole sizes,
\beq
C(E)={1\over2}\,\vec\nabla_{r_1}\!\!\!\cdot\vec\nabla_{r_2}\,\sigma_{\bar qq}(\vec r_1-\vec r_2,E)
\Biggr|_{\vec r_1=\vec r_2}.
\label{740}
\eeq
This factor steeply rises with energy. For $J/\Psi$ produced at $\sqrt{s}=7\TeV$ at the mid-rapidity $E=e^y\sqrt{s(M_{J/\Psi}^2+\la p_T^2\ra)}/2m_N=15.6\TeV\times e^y$. At this energy and $y=0$ the factor $C(E)$ was calculated in \cite{broad}
at $C(E)=13$. Inclusion of gluon shadowing corrections \cite{broad} substantially reduces this factor $C(E)\Rightarrow C_\mathrm{shad}=6.5$. We rely on this value for further evaluations.

Broadening Eq.~(\ref{700}) averaged over impact parameter is given by the the mean number of collisions, Eq.~(\ref{270}):
\beq
\Delta p_T^2 = \frac{9\,C_\mathrm{shad}}{2\,\sigma_\mathrm{in}^{pp}}\,(N_\mathrm{coll}-1)=
\frac{9\,C_\mathrm{shad}}{2\,\sigma_\mathrm{in}^{pp}\,\beta}\,\left(R_h^{pA}-1\right).
\label{780}
\eeq

\begin{figure}[htb]
\begin{center}
 \includegraphics[width=7cm]{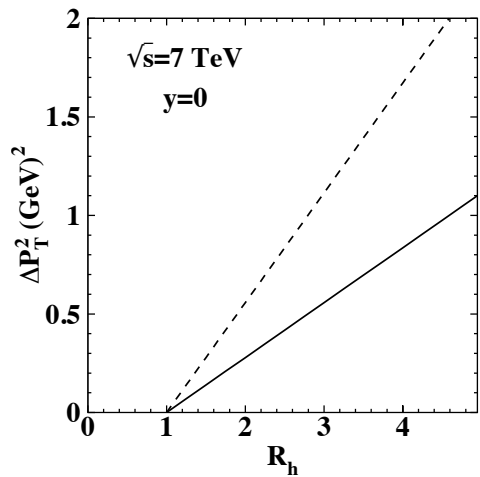} \\[-18mm]
 \caption{\label{fig:broad} $p_T$ broadening of $J/\Psi$ produced in high-multiplicity $pp$ collisions
 at $\sqrt{s}=7\TeV$ at mid-rapidity. The solid and dashed curves present broadening calculated with Eq.~(\ref{780}), including or excluding the corrections for gluon shadowing.}
  \end{center}
 \end{figure}

The expected broadening of $J/\Psi$ produced in high-multiplicity $pp$ collisions at $\sqrt{s}=7\TeV$ and $y=0$ is plotted in Fig.~\ref{fig:broad} as a solid line, as function of the normalized multiplicity $R_h^{pp}$. Calculations are done with $\beta=0.6$.
The dashed line shows, for comparison, broadening calculated without shadowing corrections.

\section{Summary}

High-multiplicity $pp$ collisions at high energies exhibit features which traditionally have been associated with nuclear effects. Here we observed a close similarity between multiple interactions in $pp$ and $pA$ collisions. In order to enhance multiple interactions in the former case one should trigger on high multiplicity of produced hadrons, while in the latter case one can reach the same multiplicity due to the increased number of collisions Eq.~(\ref{270}).  We employed the phenomenological description of the mean multiplicity in $pA$ collisions, Eq.~(\ref{300}), and the observed nuclear effects for $J/\Psi$ production, enabling us to predict the multiplicity dependence of the $J/\Psi$ production rate in $pp$ collisions.
The results agree well with the correlation between $R_{J/\Psi}^{pp}$ and $R_h^{pp}$ observed in $pp$ collisions at the LHC \cite{alice-psi-mult}. We also predicted $p_T$ broadening for $J/\Psi$ produced in high- compared with mean-multiplicity $pp$ collisions. We relied on data at mid-rapidity, since the possible rapidity dependence of $\alpha$ at the LHC energy is poorly known. If, however, the value of $\alpha$ drops at forward rapidities, this will lead to smaller values of $R_{J/\Psi}^{pA}$ at large multiplicities.
Although we employed data for $pA$ collisions integrated over impact parameter, the analysis can be done at different centralities.  

\begin{acknowledgments}

\end{acknowledgments}
This work was supported in part
by Fondecyt (Chile) grants 1130543, 1130549, 1100287, 
and by Conicyt-DFG grant No. RE 3513/1-1.


\begin{thebibliography}{99}

\bibitem{kancheli} 
O.~V.~Kancheli,
JETP Lett. 18 (1973) 274 [Pisma Zh. Eksp. Teor.
Fiz. 18 (1973) 465].

\bibitem{ak} 
  V.~A.~Abramovskii and O.~V.~Kancheli,
  Pisma Zh.\ Eksp.\ Teor.\ Fiz.\  {\bf 15}, 559 (1972).

\bibitem{karen1} 
  K.~A.~Ter-Martirosyan,
  Phys.\ Lett.\ B {\bf 44}, 377 (1973).

\bibitem{karen2} 
  A.~B.~Kaidalov and K.~A.~Ter-Martirosian,
  Phys.\ Lett.\ B {\bf 117}, 247 (1982).

\bibitem{agk} 
  V.~A.~Abramovsky, V.~N.~Gribov and O.~V.~Kancheli,
  Yad.\ Fiz.\  {\bf 18}, 595 (1973)
  [Sov.\ J.\ Nucl.\ Phys.\  {\bf 18}, 308 (1974)].

\bibitem{karen3} 
  K.~A.~Ter-Martirosyan,
  Pisma Zh.\ Eksp.\ Teor.\ Fiz.\  {\bf 15}, 734 (1972).

\bibitem{kancheli70} 
  O.~V.~Kancheli,
  JETP Lett.\  {\bf 11}, 267 (1970)
  [Pisma Zh.\ Eksp.\ Teor.\ Fiz.\  {\bf 11}, 397 (1970)].
  
\bibitem{mueller} 
  A.~H.~Mueller,
  Phys.\ Rev.\ D {\bf 2}, 2963 (1970).

\bibitem{huefner}
  W.~Q.~Chao, M.~K.~Hegab and J.~H\"ufner,
  Nucl.\ Phys.\  A {\bf 395}, 482 (1983).

\bibitem{froissaron} 
  M.~S.~Dubovikov, B.~Z.~Kopeliovich, L.~I.~Lapidus and K.~A.~Ter-Martirosian,
  Nucl.\ Phys.\ B {\bf 123}, 147 (1977).

\bibitem{qgsm} A.B. Kaidalov, JETP Lett. {\bf 32}, 474 (1980) [Sov. J. Nucl. Phys. {\bf 33}, 733 (1981)]; 
Phys. Lett. B{\bf 116}, 459 (1982).

\bibitem{capella} A. Capella et al., Phys. Rep. 236 (1994) 225.

\bibitem{kaidalov} 
  A.~B.~Kaidalov,
  Phys.\ Lett.\ B {\bf 116}, 459 (1982).

\bibitem{kaidalov-karen} 
  A.~B.~Kaidalov and K.~A.~Ter-Martirosian,
  Sov.\ J.\ Nucl.\ Phys.\  {\bf 39}, 979 (1984)
  [Yad.\ Fiz.\  {\bf 39}, 1545 (1984)];
  Sov.\ J.\ Nucl.\ Phys.\  {\bf 40}, 135 (1984)
  [Yad.\ Fiz.\  {\bf 40}, 211 (1984)].

\bibitem{glr} L.V.~Gribov, E.M.~Levin and M.G.~Ryskin, Nucl. Phys. {\bf
B188} (1981) 555; Phys. Rep. {\bf 100}, 1 (1983).

\bibitem{al}
  A.~H.~Mueller,
  arXiv:hep-ph/9911289.

\bibitem{mv}
  L.~D.~McLerran and R.~Venugopalan,
  Phys.\ Rev.\  D {\bf 49}, 2233 (1994);
  Phys.\ Rev.\  D {\bf 49}, 3352 (1994);
  Phys.\ Rev.\  D {\bf 50}, 2225 (1994).

\bibitem{busza} 
  J.~E.~Elias, W.~Busza, C.~Halliwell, D.~Luckey, P.~Swartz, L.~Votta and C.~Young,
  Phys.\ Rev.\ D {\bf 22}, 13 (1980).

\bibitem{phobos} 
  B.~B.~Back {\it et al.}  [PHOBOS Collaboration],
  Phys.\ Rev.\ C {\bf 72}, 031901 (2005).

\bibitem{alice-mult} B. Abelev et al. (ALICE Collaboration),
Phys. Rev. Lett. 110, 032301 (2013).

\bibitem{mine} 
  B.~Z.~Kopeliovich,
  Phys.\ Rev.\ C {\bf 68}, 044906 (2003).
  
\bibitem{kps-heraB} 
  B.~Z.~Kopeliovich, I.~K.~Potashnikova and I.~Schmidt,
  Phys.\ Rev.\ C {\bf 73}, 034901 (2006).  
  
  \bibitem{ciofi} 
  C.~Ciofi degli Atti, B.~Z.~Kopeliovich, C.~B.~Mezzetti, I.~K.~Potashnikova and I.~Schmidt,
  Phys.\ Rev.\ C {\bf 84}, 025205 (2011).

\bibitem{na3} 
  J.~Badier {\it et al.}  [NA3 Collaboration],
  Z.\ Phys.\ C {\bf 20}, 101 (1983).

\bibitem{puzzles} 
  B.~Z.~Kopeliovich,
  Nucl.\ Phys.\ A {\bf 854}, 187 (2011).

\bibitem{e866}
  M.~J.~Leitch {\it et al.}  [FNAL E866 Coll.],
  Phys.\ Rev.\ Lett.\  {\bf 84}, 3256 (2000).
  
\bibitem{phenix-psi} 
  A.~Adare {\it et al.}  [PHENIX Collaboration],
  Phys.\ Rev.\ Lett.\  {\bf 107}, 142301 (2011).
  
 \bibitem{alice-psi} R. Arnaldi [ALICE Coll.] CERN LHC Seminar, June 18th 2013.

\bibitem{boosting} 
  B.~Z.~Kopeliovich, H.~J.~Pirner, I.~K.~Potashnikova and I.~Schmidt,
  Phys.\ Lett.\ B {\bf 697}, 333 (2011).

\bibitem{cap-kop} 
  A.~Capella and B.~Z.~Kopeliovich,
  Phys.\ Lett.\ B {\bf 381}, 325 (1996).

\bibitem{koplik} 
  J.~Koplik and A.~H.~Mueller,
  Phys.\ Rev.\ D {\bf 12}, 3638 (1975).

\bibitem{spots} 
  B.~Z.~Kopeliovich, I.~K.~Potashnikova, B.~Povh and I.~Schmidt,
  Phys.\ Rev.\ D {\bf 76}, 094020 (2007).

\bibitem{bkk} 
  K.~G.~Boreskov, A.~B.~Kaidalov and O.~V.~Kancheli,
  Phys.\ Atom.\ Nucl.\  {\bf 69}, 1765 (2006)
  [Yad.\ Fiz.\  {\bf 69}, 1802 (2006)].

\bibitem{morsch} 
  B.~Abelev {\it et al.}  [ALICE Collaboration],
  arXiv:1307.1094 [nucl-ex].

\bibitem{pisa} A.~DiGiacomo and H.~Panagopoulos, Phys.  Lett. B {\bf 285}, 133
(1992).

\bibitem{shuryak} E. Shuryak and I.~Zahed, Phys. Rev. D69 (2004) 014011.

  \bibitem{kst2} B.~Z.~Kopeliovich, A.~Sch\"afer and A.~V.~Tarasov,
  Phys.\ Rev.\  D {\bf 62}, 054022 (2000).
  
\bibitem{k3p} 
  B.~Z.~Kopeliovich, I.~K.~Potashnikova, B.~Povh and E.~Predazzi,
  Phys.\ Rev.\ Lett.\  {\bf 85}, 507 (2000);
  Phys.\ Rev.\ D {\bf 63}, 054001 (2001).
  
\bibitem{totem} 
  B.~Z.~Kopeliovich, I.~K.~Potashnikova and B.~Povh,
  Phys.\ Rev.\ D {\bf 86}, 051502 (2012).

 \bibitem{alice-psi-mult} 
  B.~Abelev {\it et al.}  [ALICE Collaboration],
  Phys.\ Lett.\ B {\bf 712}, 165 (2012)

\bibitem{jkt}
  M.~B.~Johnson, B.~Z.~Kopeliovich and A.~V.~Tarasov,
  Phys.\ Rev.\  C {\bf 63}, 035203 (2001).

\bibitem{broad} 
  B.~Z.~Kopeliovich, I.~K.~Potashnikova and I.~Schmidt,
  Phys.\ Rev.\ C {\bf 81}, 035204 (2010).
  
 \bibitem{dhk}
J.~Dolejsi, J.~H\"ufner and B.~Z.~Kopeliovich,
  Phys.\ Lett.\  B {\bf 312}, 235 (1993).

\end{thebibliography}
\end{document}